\documentclass[12pt,floats]{article}

\usepackage[]{umlaut,latexsym}
\usepackage[xdvi]{graphicx}

\usepackage{natbib}
\usepackage{amssymb}

\begin{document}

\parskip 2mm

\renewcommand{\refname}{\normalsize \bf \em References}

\title{\bf EXCHANGE INTERACTIONS AND MAGNETIC ANISOTROPY 
           IN THE ``Ni$_4$'' MAGNETIC MOLECULE } 
%
%
\author{
      A.\ V.\ POSTNIKOV,$^{a,b\,}$\footnote{Corresponding author:
      Tel.: +49 541 969 2377,
      fax:  +49 541 969 2351,
      email: \mbox{apostnik@uos.de}}~
      M.\ BR\"UGER$^a$
      and J.\ SCHNACK$^a$
\\*[0.2cm]
      $^a${\small \it Universit\"at Osnabr\"uck -- Fachbereich Physik,
      D-49069 Osnabr\"uck, Germany;} \\
      $^b${\small \it Institute of Metal Physics,
      S. Kowalewskoj 18, Yekaterinburg 620219, Russia}
}
\maketitle

\begin{abstract}
%
Magnetic properties of a tetrahedral ``Ni$_4$'' molecule
are discussed in terms of the Heisenberg model, with magnetic anisotropy
terms included, and on the basis of first-principle calculations
within the density functional theory. It is shown that the isotropic
Heisenberg model does not provide an adequate description of
magnetization at low temperatures; an inclusion of single-site
anisotropy terms does not help to improve the situation either. We suppose
that the magnetostriction of the molecule and hence the dependence of
interatomic coupling parameters on the magnetization might be important
for an adequate description of magnetic properties.
The first-principle calculations confirm the system's general preference
for antiferromagnetic coupling, as well as the failure of the isotropic
Heisenberg model.
A conjugated-gradient search for the relaxed structure of the ``Ni$_4$'' 
molecule in ferromagnetic and antiferromagnetic configurations did not show
any clear tendency to diversification of interatomic distances.
These calculations however have not yet included the spin-orbit coupling,
which can be essential for analyzing the effects of magnetostriction.
%
\end{abstract}


\section*{1. INTRODUCTION}

The system of our present interest belongs to a large family of
molecular magnets --  metallo-organic or -anorganic substances 
in which well defined
molecular fragments possess intrinsic magnetic structure, and can
crystallize forming a stable sold phase. An introduction into the subject 
can be found in the books by \citet{Kahn-book}, or \citet{Mol_Magnets}.%
\footnote{In a broader perspective,
one includes sometimes purely organic magnetic substances, on one side,
and three-dimensional connected metalloorganic systems, on the other side,
in the definition of molecular magnets.} 
The practical interest for molecular magnets is heated by high net spin moments
and/or large magnetic anisotropy in some of them, that seems promising 
for dense magnetic storage. Moreover, unique and novel properties 
of molecular magnets, like quantum tunneling of magnetization
\citep{PRL76-3830,PRL82-3903}, open perspectives of basically new
applications -- manipulation of magnetic states by light, work media
for quantum computers, etc.

\noindent
In the actual stage of ``extensive'' exploration of the field, 
as clear guidelines relating chemical composition to properties
are still missing, one witnesses a permanent quest on the side of 
chemists to synthesize new systems with unusual or otherwise so far unknown 
combination of metal or organic building blocks.
``Ni$_4$'' is such a system, where the magnetic coupling between
four Ni ions (each carrying spin $s$=1) occurs via a long and
chemically not very common path.

\noindent
In the present contribution, we outline the analysis of known
magnetic properties in terms of the Heisenberg model in its simplest
form and also with some extensions. Simultaneously, we perform
first-principles simulations of the electronic structure and
try to estimate magnetic interaction parameters from them.

\section*{2. STRUCTURE AND BASIC MAGNETIC \newline 
\hspace*{0.8cm}PROPERTIES}

\begin{figure}[t!] 
  \begin{center}
  \includegraphics*[width=12cm]{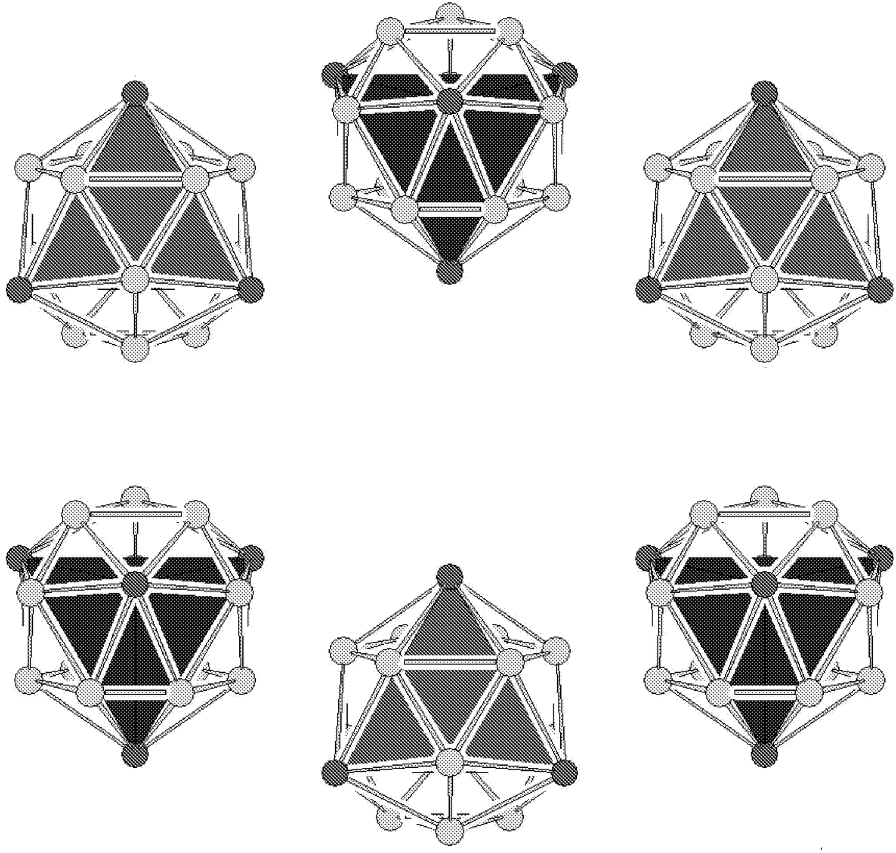}
  \end{center}
\medskip
\caption{\small
The arrangement of basic structural units Ni$_4$Mo$_{12}$ units in
the molecular crystal (cortesy of Paul K\"ogeler). A slightly distorted 
Ni$_4$ tetrahedron (dark circles) is kept together by O-bridged Mo$_{12}$ 
cage (gray circles). 
\label{fig:struc1}
}
\end{figure}

The full chemical formula of ``Ni$_4$'' is
[Mo$_{12}$O$_{30}$($\mu_2$-OH)$_{10}$H$_2$\{Ni(H$_2$O)$_3$\}$_4$]
$\cdot$14 H$_2$O;
its synthesis and characterization have been reported 
by \citet{InCh39-5176}.
The basic structural unit is a slightly distorted Ni$_4$ tetrahedron,
whose vertices cap the hexagonal faces of the O-bridged Mo$_{12}$ cage. 
The crystal structure contains 
two such units, related by the 180$^{\circ}$ rotation around an edge 
of the tetrahedron, as is shown in Fig.~\ref{fig:struc1}.
The nearest neighbourhood of each Ni atom is (nearly) octahedral O$_6$,
with three apical oxygens saturated by bonded hydrogen, and three
others participating in the bonding to Mo atoms.
The magnetic interaction path is therefore Ni--O--Mo$_2$--O--Ni,
with the length of 6.6--6.7 {\AA}. 
The nominal Ni$^{I\!I}$ state corresponds to the $3d^8$ configuration 
and hence $s$=1. The ground state was found to have the total spin $S$=0
\citep{Schnack-Ni4}.

\noindent
The quantitative description of magnetic properties (magnetization and
magnetic susceptibility) reported by \citet{InCh39-5176} requires
to choose an underlying physical model. The traditional simplest
choice is that of the Heisenberg model, which we introduce 
(including the Zeeman term) as follows:
%
\begin{eqnarray}
%
{\cal H} = J\!\sum_{\{\alpha,\beta\} \atop \mbox{\tiny pairs}}^6\!
\mathbf{s}_{\alpha}\!\cdot\!\mathbf{s}_{\beta}
+ g\mu_{\mbox{\tiny B}}
\sum_{\alpha}^4\,\mathbf{B}\!\cdot\!\mathbf{s}_{\alpha}\,.
\label{eq:Heis}
\end{eqnarray}
\noindent
All $s_{\alpha}$=1, and $J>0$ corresponds to antiferromagnetic (AFM) 
coupling in this formulation.
In the following, we would need to extend this model over anisotropic effects.
The simplest case is an inclusion of the single-ion anisotropy,
%
\begin{eqnarray}
%
{\cal H} = J\!\sum_{\{\alpha,\beta\}\atop \mbox{\tiny pairs}}^6\!
\mathbf{s}_{\alpha}\cdot\mathbf{s}_{\beta}
+ D\!\left[\sum_{\alpha}^4(\mathbf{e}_{\alpha}\!\cdot\!\mathbf{s}_{\alpha})^2
- \frac{8}{3}\right]
+ g \mu_{\mbox{\tiny B}}\sum_{\alpha}^4\,\mathbf{B}\cdot\mathbf{s}_{\alpha}\,.
\label{eq:aniso}
\end{eqnarray}
\noindent
with $D$ being the constant of zero-field splitting, and ${\vec e}_{\alpha}$
the local anisotropy axis for each ion.

\begin{figure}[t!] 
  \begin{center}
  \includegraphics*[width=13cm]{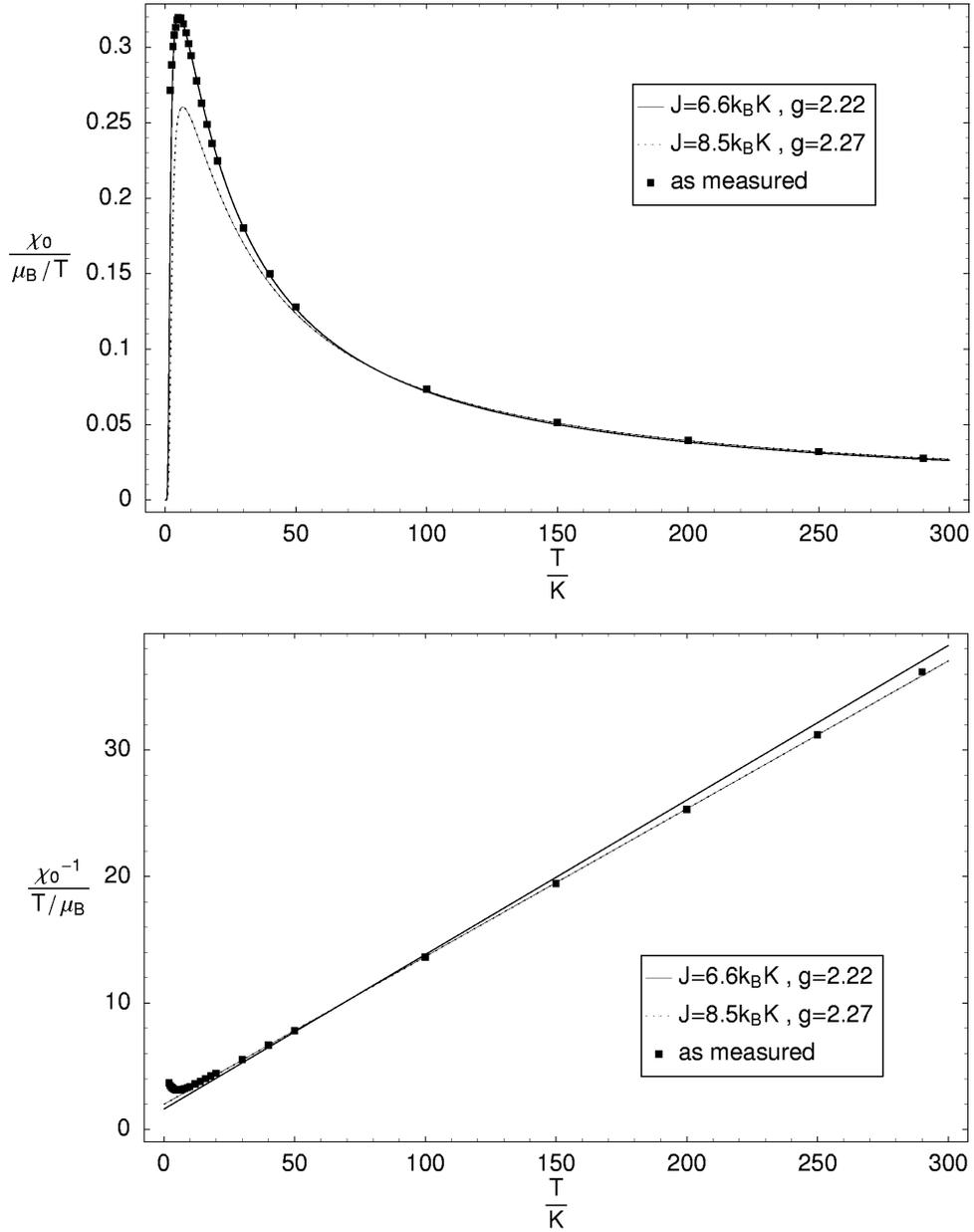}
  \end{center}
\medskip
\caption{\small
Magnetic susceptibility (top panel) and its inverse (bottom panel), 
as reported by \citet{InCh39-5176},
in comparison to the fit of the Curie--Weiss law with two sets of
parameters \citep{Brueger-diplom}.
\label{fig:suscep}
}
\end{figure}

\noindent
With $N$=4 coupled spins $s$=1, the total dimension of problem is
%
\begin{eqnarray}
(2s+1)^N = 81 = \sum_{S=0}^4\;\sum_{M_S=-S}^S G_S\,,
\label{eq:H-dimen}
\end{eqnarray}
\noindent
where $G_{S=0,1,2,3,4} = 3,6,6,3,1$ are degeneracies of 
the $|S, M_S\rangle$ states.
The exact diagonalisation of this model 
can be straightforwardly done for any choice of parameter values. The magnetic 
susceptibility $\chi$ and magnetization $M$ are obtained 
in terms of the partition function $Z$:
%
\begin{eqnarray}
Z &=& \sum_{S=0}^4 \sum_{M=-S}^S G_S\;
e^{-\frac{\displaystyle E_{S,M_S}}{\displaystyle k_{\mbox{\tiny B}}T}}\,;\\
M &=& -\frac{g\mu_{\mbox{\tiny B}}}{Z}\;\mbox{Tr}\left[
e^{-\frac{\displaystyle \cal H}{\displaystyle k_{\mbox{\tiny B}}T}}
S_Z\right]\,,\\
\chi &=& \frac{\partial M}{\partial B} = 
\frac{(g\mu_{\mbox{\tiny B}})^2}{k_{\mbox{\tiny B}}T\,Z}\;
\mbox{Tr}\left[
e^{-\frac{\displaystyle \cal H}{\displaystyle k_{\mbox{\tiny B}}T}}
S_Z^{\,2}\right]\ - 
\frac{M^2}{k_{\mbox{\tiny B}}T}\,. 
\end{eqnarray}
\noindent
Fig.~\ref{fig:suscep} shows the fit of the measured magnetic susceptibility
data, as obtained by Paul K\"ogerler on powder samples and reported
by \citet{InCh39-5176}, to the isotropic Heisenberg model of
Eq.~(\ref{eq:Heis}). The Heisenberg model yields the Curie--Weiss
behaviour,
%
\begin{eqnarray}
%
\chi_{\mbox{\tiny CW}}=\frac{C}{T+T_{\mbox{\tiny C}}}
\quad\mbox{with}\quad
C=\frac{8(g\mu_{\mbox{\tiny B}})^2}{3k_{\mbox{\tiny B}}}
\quad\mbox{and}\quad
T_{\mbox{\tiny C}}=\frac{2J}{k_{\mbox{\tiny B}}}\,,
\label{eq:suscep}
\end{eqnarray}
\noindent
from fitting to which one can estimate the values of $J$ and $g$.
Fig.~\ref{fig:suscep} shows that both direct and inverse susceptibility
can be satisfactorily fitted through wide temperature range 
with only slightly varying parameter values, either 
$\left\{J\mbox{=6.6 K}; g\mbox{=2.22}\right\}$, or 
$\left\{J\mbox{=8.5 K}; g\mbox{=2.27}\right\}$
\citep{Brueger-diplom}.

\begin{figure}[t!]
  \begin{center}
  \includegraphics*[width=13cm]{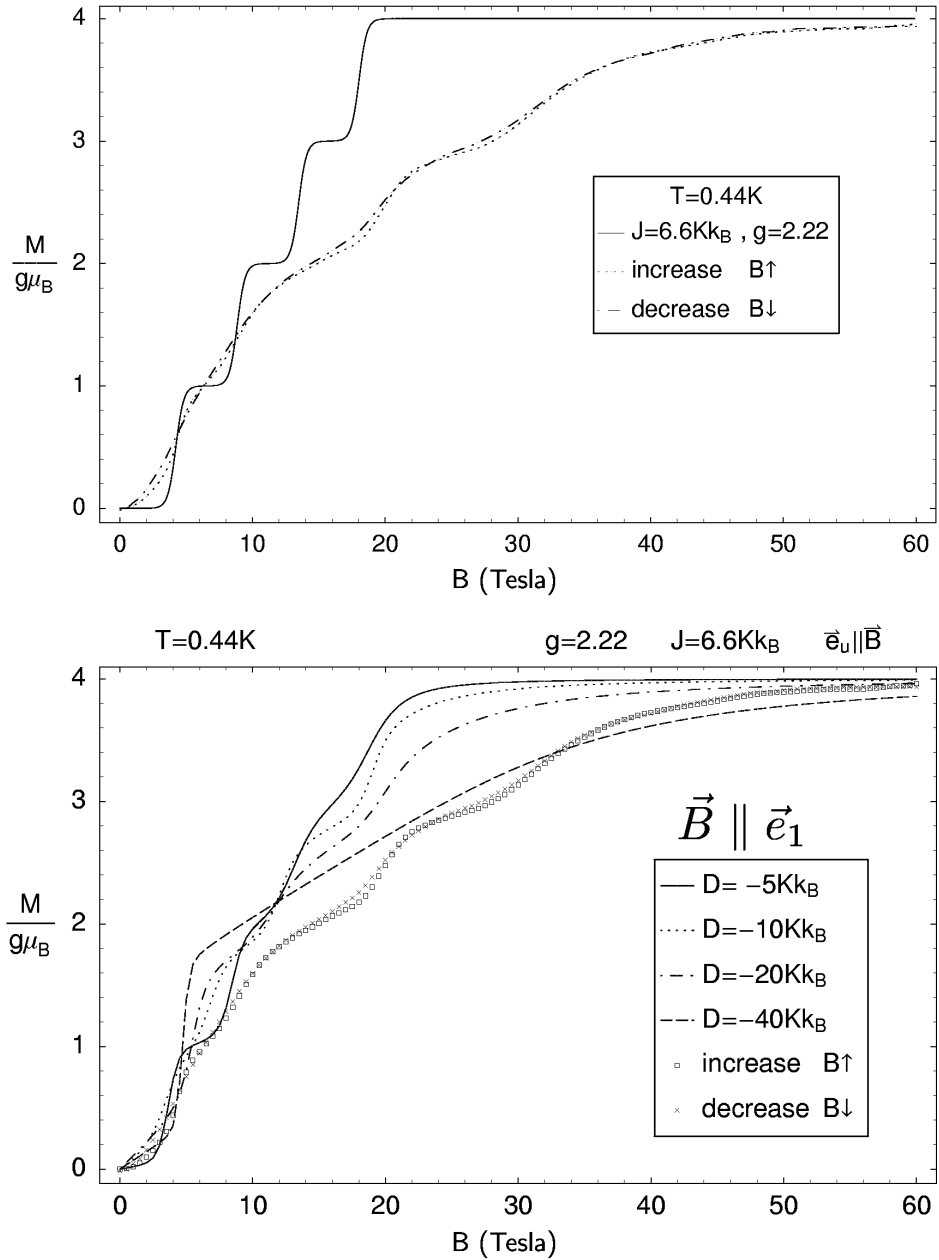}
  \end{center}
\medskip
\caption{\small
An attempted fit \citep{Brueger-diplom} to the measured 
\citep[by][]{InCh39-5176} magnetization as function of the external 
magnetic field: within the isotropic Heisenberg model (top panel) 
and allowing different values of the zero-field splitting parameter $D$ 
(bottom panel). 
\label{fig:magnet}
}
\end{figure}

\begin{figure}[t!] 
  \begin{center}
  \includegraphics*[width=13cm]{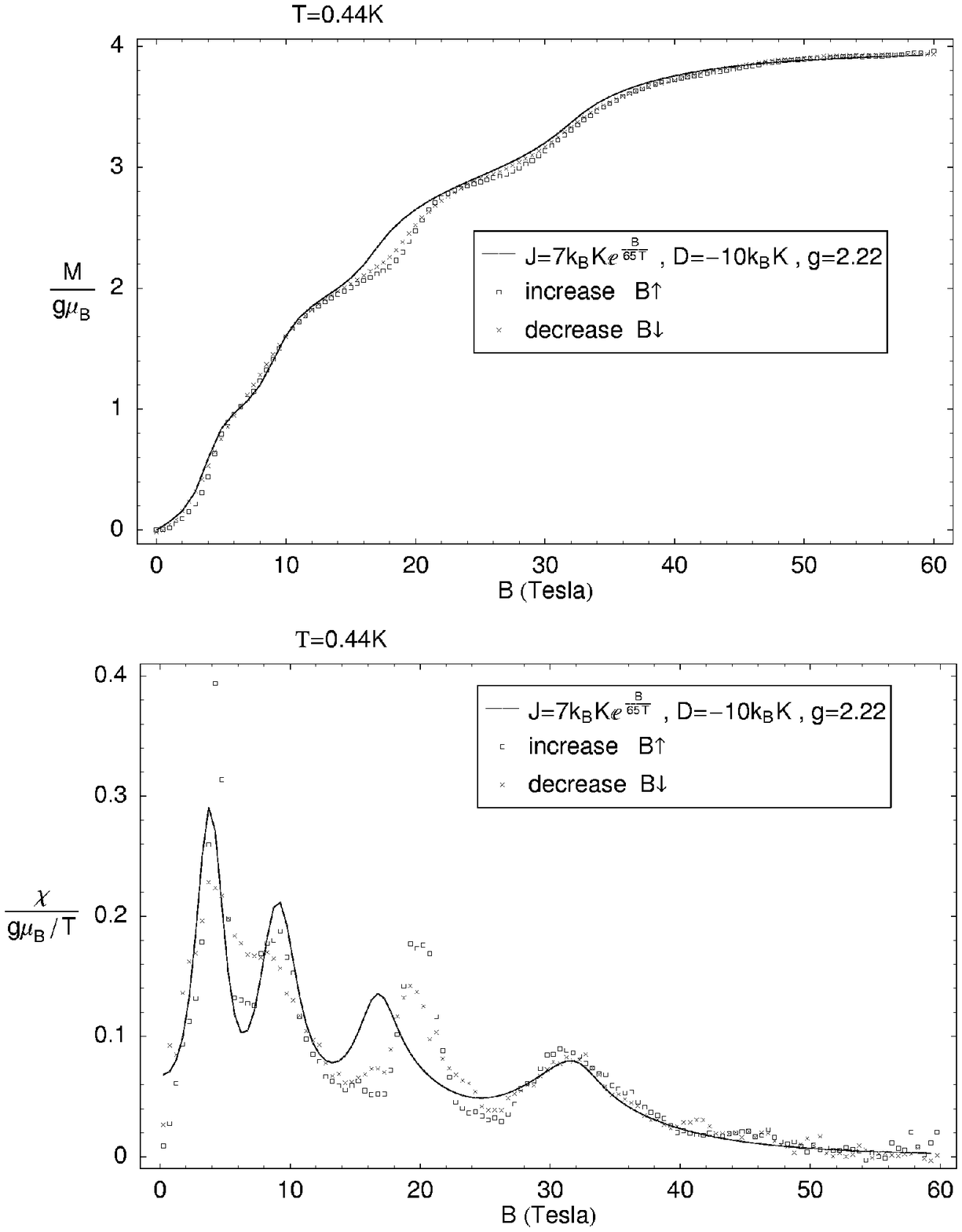}
  \end{center}
\medskip
\caption{
\small
A fit to the measured 
\citep[by][]{InCh39-5176} magnetization as function of the external 
magnetic field $B$, according to the Heisenberg model with
zero-field splitting $D$, averaged over 1000 randomly generated
orientations, in combination with $B$-dependent magnetic coupling (top panel);
the corresponding magnetic susceptibility
(bottom panel). 
}
\label{fig:fitJB}
\end{figure}

\noindent
However, the observed field dependence of the magnetization,
measured at sufficiently low temperature (0.44 K), cannot be
reasonably described by any of these sets of parameters, nor by any
other one, within the isotropic Heisenberg model.  
As is seen in Fig.~\ref{fig:magnet}, the Heisenberg model yields
equidistant steps in the magnetization curve, with appropriate 
temperature smearing.
In the experimental curve, on the contrary, the steps are not 
equidistant (in the field strength), moreover they are much stronger
smeared, to the extend that the first step is hardly pronounced.

\noindent
A number of attempts has been undertaken by \citet{Brueger-diplom}
to achieve a reasonable fit to the measured magnetization curve,
varying the values of magnetic coupling parameter $J$ and
of the zero-field splitting parameter $D$, various
orientations of the external magnetic field relative to the Ni$_4$
tetrahedron, averaging over different orientations, etc.
A non-equidistant steps in the magnetization, somehow similar to
those experimentally observed, can be simulated by allowing an increase
of coupling constants $J$ with magnetic field. As the physical
mechanisms of interatomic coupling, in the simplest picture, have to do 
with the overlap of orbitals centered at different sites,
a justification of the $J(B)$ dependency might be provided by the 
variation of interatomic distances in the magnetic field, i.e.,
the magnetostriction effects. The exponential dependence of interatomic
overlap integrals with distance would justify the parametrization
$J(B) \sim \exp(B/\gamma)$.

\noindent
Fig.~\ref{fig:fitJB} shows a fit to the experimental magnetization 
and magnetic susceptibility with a zero-field splitting constant included
in the Heisenberg model \emph{and} an empirical magnetic field dependency
of the magnetic interaction parameter $J$. The exact diagonalisation
of the corresponding model Hamiltonian yields an almost
acceptable description of the experimental trends, yet leaves
the questions open about the microscopic origins, and the specific form, 
of the model parameters' variation with magnetic field.

\section*{3. DFT CALCULATIONS SETUP}

The first-principles approach, on the contrary, is able to compare
the energies of different magnetic configurations without making
assumptions about relevant physical interactions in the system.
Moreover, the values of model Hamiltonian parameters, like exchange
couplings $J$ and anisotropy constants $D$, can be in principle
evaluated without referring to any experimental data. An abundant literature
exists notably to the subject of $J$ \citep[see, e.g.,]%
[for a recent discussion on the subject]{JCP116-2728,JCP116-3985}.
We emphasize the difference between multi-configurational quantum-chemical
approach, and that of the density functional theory (DFT).
The former refers to the differences in energy of, 
e.g., singlet and triplet multi-determinantal states.
In the DFT, such properties are not accessible, but one can compare
the expectation values of the Hamiltonian in specially prepared states,
e.g., a triplet (ferromagnetic) and the broken-symmetry
(antiferromagnetic) ones. For the comparison of approaches and a more
detailed discussion, see \citet{Psik-highlight}.
Further reference magnetic configurations can be created in the DFT
by using the fixed spin moment (FSM) scheme \citep{JPF14-L129}.
The discussion on the applicability of the Heisenberg model
based on the results of DFT calculations and comparison of the FSM 
results for another molecular magnet, hexanuclear ``ferric wheel'',
has been given in \citet{EMRS-Fewheel}. 
In the following, we make a similar analysis for the ``Ni$_4$'' system.
As in the last cited paper, the DFT calculations have been done with the
{\sc Siesta} method \citep{JPCM14-2745}, using norm-conserving 
pseudopotentials \citep{PRB43-1993}
and a compact basis set of strictly localized numerical functions.
Specifically, we used a triple-$\zeta$ basis for Mo$4d$ and Ni$3d$ states,
double-$\zeta$ with polarization orbitals for O$2s$ and $2p$, H$1s$, Mo$4p$, 
Ni$4p$, and single-$\zeta$ with polarization orbitals for Mo$5s$ and
Ni$5p$. The energy shift parameter, governing the spatial confinement of
basis function, was 20 mRy. The exchange-correlation was treated in
the generalized gradient approximation after \citet{PRL77-3865}.
A molecular fragment with 104 atoms, in total, was treated as an
isolated molecule in a cubic simulation box with the edge size 20 {\AA}. 
The cutoff parameter defining the real-space mesh for solving
the Poisson equation was set to 200 Ry, that resulted in a
180$\times$180$\times$180 grid in the abovementioned cubic simulation cell.
In order to smoothen the variations of the total energy and other properties
due to the ``eggbox effect'', an averaging of calculated grid-dependent
properties over a finer fcc-type sub-grid has been done, as described
by \citet{JPCM14-2745}.
The calculations have been done for the ``as determined'' crystal
structure, i.e., corresponding to the ground state of $S$=0.
We allowed moreover a structure relaxation for antiferromagnetic
(AFM, $S$=0) and ferromagnetic (FM, $S$=1) configurations.

\noindent
Before discussing the results it should be noted that the {\sc Siesta}
method does not include the treatment of spin-orbit interaction,
hence it is not possible to discuss the magnetic anisotropy parameters
on the basis of calculation. In principle, the calculation of
zero-field splitting $D$ is not only possible, but yields the quantitative
results of very high accuracy. \citet{PRB59-R693,PRB60-9566} 
proposed a method for calculation of the second-order anisotropy energies,
which yielded perfect agreement with experiment for the Mn$_{12}$-acetate
(see the cited works) and for a number of other molecular magnets
\citep[see Table 3 in][]{Psik-highlight}.

\section*{4. FIXED SPIN MOMENT CALCULATIONS}

The FSM scheme fixes the total number of electrons in majority- and 
minority-spin channels. This amounts to imposing an effective external 
magnetic field, which -- at least in metal systems -- introduces
a difference in the chemical potential for majority- and minority-spin
electrons. When applied to molecular magnets, the fixed total moment
can be only an integer (and usually even) amount of Bohr magnetons.
Isolated magnetic molecules have discrete Kohn-Sham energy levels,
and normally, a non-zero HOMO-LUMO gap. Therefore both spin channels may
still possess a common chemical potential (if their respective HOMO-LUMO 
gaps do overlap), or they may need, similarly to the case of a metallic 
system, an external magnetic field for enforcing the splitting. 
In the former case, the DFT solutions corresponding to different
FSM values coexist as metastable states, whose total energies can be
directly compared. In the latter case, a Zeeman term must be taken
into account as an additional energy needed to split the chemical
potentials in two spin channels. These are the foundations of our
FSM analysis for Ni$_4$. 

\begin{figure}[t!] 
  \begin{center}
  \includegraphics*[width=14.5cm]{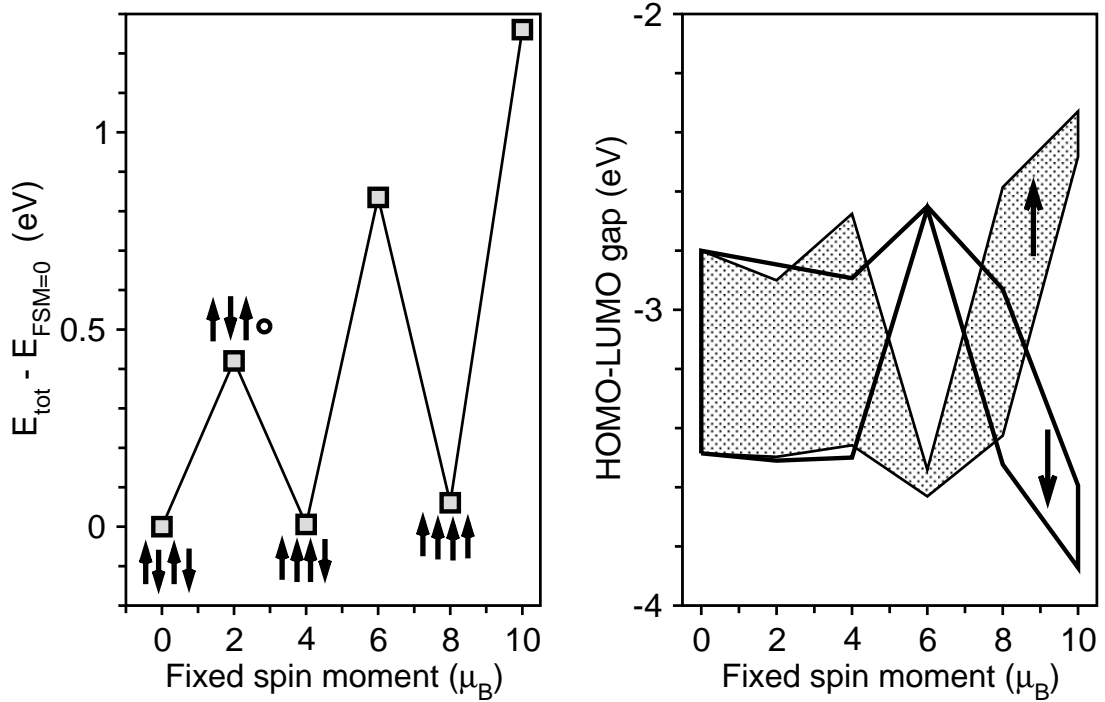}
  \end{center}
\medskip
\caption{\small
Total energy relative to that of the AFM ground state (left panel)
and HOMO-LUMO gap values in two spin channels (right panel)
from the FSM calculations. The relative orientation 
of magnetic moments associated with Ni atoms
(nominally $s$=1) is shown schematically in the left panel 
for different magnetic configurations.
\label{fig:fsm}
}
\end{figure}

\begin{figure}[p!] 
  \begin{center}
  \includegraphics*[width=14cm]{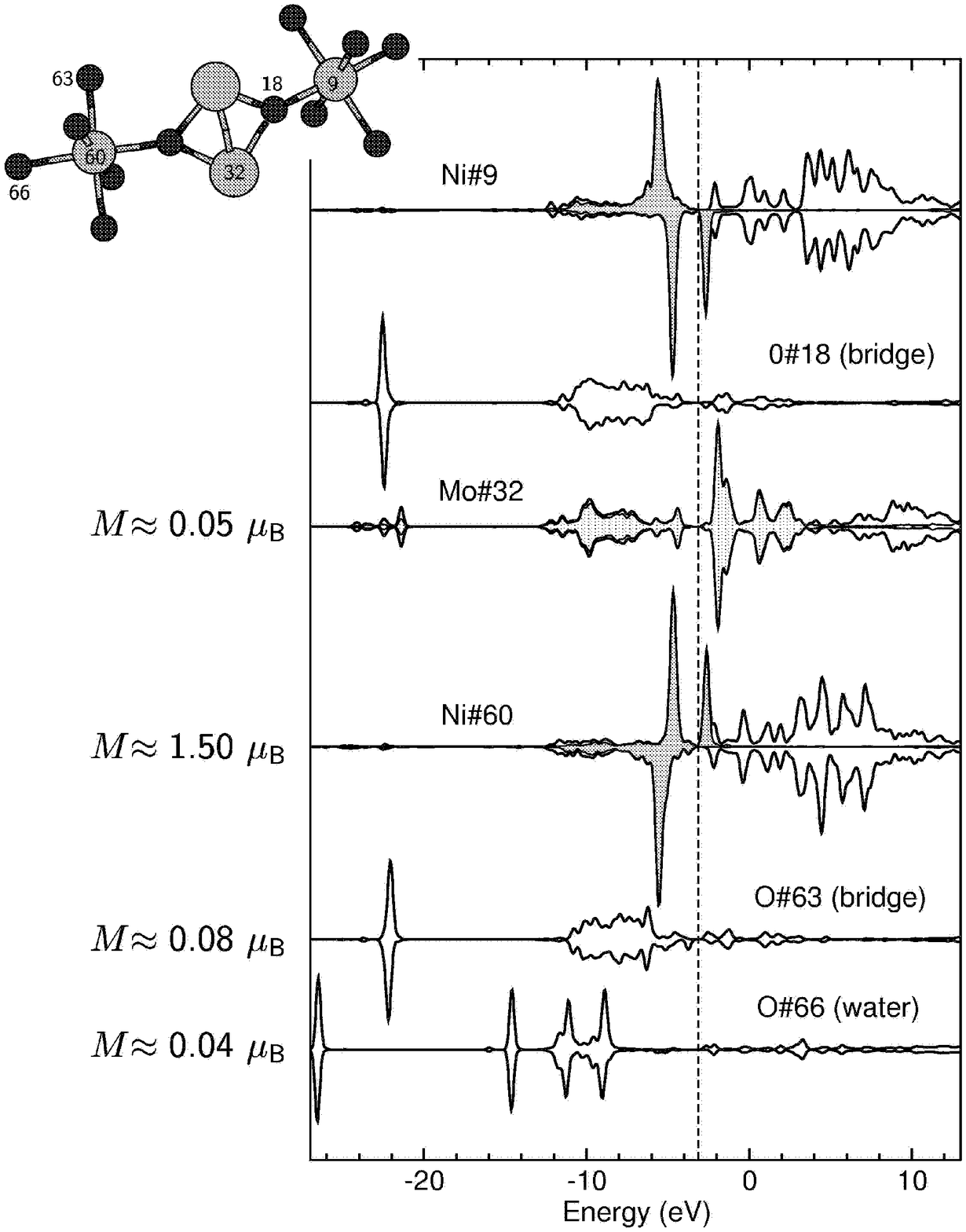}
  \end{center}
\medskip
\caption{\small
Local spin-resolved DOS 
of two Ni atoms in AFM configuration along with
bridging O and Mo atoms. The inset shows the corresponding
part of the ``Ni$_4$'' molecule with the numbering of atoms.
Local magnetic moments according to Mulliken population
analysis are indicated on the left. The shaded areas for Ni (Mo) atoms
show the contributions of the $3d$ ($4d$) states.
\label{fig:dos}
}
\end{figure}

\noindent
Fig.~\ref{fig:fsm} shows the total energies for a number of configurations
with different FSM values (left panel), along with corresponding
HOMO-LUMO gaps in two spin channels. For setting up an AFM configuration
(zero FSM value), the local magnetic moments of two (arbitrary, due to
nearly perfect tetrahedral symmetry) Ni atoms were inverted. This
configuration then always survived in the course of self-consistent
calculation.
The corresponding spin-resolved local densities of states (DOS) for
two oppositely magnetized Ni atoms and several of their neighbours are shown
in Fig.~\ref{fig:dos}. A fragment of the molecule, including the path
between two Ni atoms, is depicted in the inset. It is well seen
that the local DOS at the Ni site in the majority-spin channel
is dominated by a fully occupied $3d$ subband, which forms a single
narrow peak. In the minority-spin channel, one sees a crystal-field
splitting into the $t_{2g}$ (occupied) and the $e_g$ (empty) energy levels.
Such clear separation, without a pronounced further splitting, 
is due to the fact that the oxygen octahedra around the Ni atoms
are almost undistorted. The above electronic configuration corresponds
to the spin value $s$=1 per Ni atom, consistently with experiment.
The value of the \emph{local} magnetic moment per Ni atom, as determined
by the Mulliken population analysis, may however differ from
2 $\mu_{\mbox{\tiny B}}$, depending on the spatial distribution of
the spin density and its decomposition into different basis orbitals.
In particular, the charge transfer from Ni to O reduces the nominal value
of the majority-spin Ni$3d$ occupation, at the same time inducing
magnetic moments at the oxygen ligands. The values of local magnetic moments
at several representative atoms is shown in Fig.~\ref{fig:dos}
on the left to each plot. Obviously the Ni atom with its six O neighbours 
makes a fragment which essentially carries the magnetic moment 
of 2 $\mu_{\mbox{\tiny B}}$; the magnetization
of Mo atoms is not pronounced.

\noindent
Such cumulative spins $s$=1 per NiO$_6$ fragment remain relatively intact
if one imposes different values of the fixed spin moment. The AFM state
(FSM = 0 $\mu_{\mbox{\tiny B}}$), the one with the spin of one NiO$_6$ group
inverted (FSM = 4 $\mu_{\mbox{\tiny B}}$) and the FM state 
(FSM = 8 $\mu_{\mbox{\tiny B}}$) can all easily be realized, and differ
only slightly in their total energies. It is noteworthy that the imposed
FSM value of 2 $\mu_{\mbox{\tiny B}}$ converges to a configuration where
three NiO$_6$ groups maintain the spins $s$=1, and the fourth Ni atom
becomes non-magnetic (see Fig.~\ref{fig:fsm}, left panel). This configuration
lies higher in energy than the three previously mentioned ones.
Yet in all four cases there is a common gap between occupied and
unoccupied states in both spin channels (see Fig.~\ref{fig:fsm}, right panel),
that means that all four configurations can be realized as metastable
magnetic solutions, without imposing an external magnetic field and hence
without an extra energy cost due to the Zeeman term. 

\noindent
The energy of the FSM = 2 $\mu_{\mbox{\tiny B}}$ configuration cannot be
used for the mapping of calculation results to the Heisenberg model,
because it contains a different number of isolated spins as the other three
cases. For FSM =0, 4, and 8 $\mu_{\mbox{\tiny B}}$ the difference 
in energies comes from different orientation of ``rigid'' magnetic moments,
and hence can serve for an attempt to extract the magnetic interaction 
parameters. As can be seen in Table I, which compares the energy
differences between three orientations of four $s$=1 spins
according to the Heisenberg model and to the calculation results,
such extraction is not straightforward. Actually it is an indication
that the Heisenberg model does not reasonably describe magnetic
interactions in the ``Ni$_4$'' molecule. This qualitative observation
is consistent with the analysis of experimental data discussed above.
The best one can do is to provide an order-of-magnitude estimate
of the Heisenberg exchange parameter, which is positive, i.e.
favouring the antiferromagnetic coupling (in the definition of
Eq.~\ref{eq:Heis}), and hence consistent with experiment,
and has the magnitude $J \approx$ 35--87 K. The earlier discussed
$J$ value estimated from the fit to experimental results of
magnetic susceptibility (Fig.~\ref{fig:suscep}) are 6--9 K. 
To some extent, the difference can be due to underestimating
the intraatomic correlation effects in a conventional DFT calculation. 
It has been shown by \citet{PRB65-184435} for the Mn$_{12}$-acetate molecule
and by \citet{PRB67-134408} for the ``V$_{15}$'' molecular magnet
that an artificial enhancement of the on-site Hubbard-like correlation
within the LDA+$U$ formalism \citep{LDA+U}
favours the localization of the magnetic moment at the $3d$ atoms
and tends to reduce the interatomic magnetic interactions. 
However, the selection of ``correlated'' states has to be done
\emph{ad hoc} in the LDA+$U$ formalism, and the magnitude of the necessary 
correction enters the calculation as an external tuning parameter.

\begin{table}[t!]
\caption{\small
An attempted mapping of the calculated total energies 
for three metastable FSM solutions onto the Heisenberg model.
The relations between total energy values are not compatible
with the predictions of the model. Order-of-magnitude estimates 
of the $J$ parameter from two energy differences to the ground state 
are 87 K and 35 K.
}
\begin{center}
\begin{tabular*}{\textwidth}{c@{\extracolsep\fill}c@{\extracolsep\fill}c}
\hline
FSM ($\mu_{\tiny B}$) \rule[-6mm]{0mm}{12mm} & 
$\sum\limits_{\{i,j\}\mbox{\footnotesize pairs}}\!\!\!J\,
 \mathbf{s}_i\,\mathbf{s}_j$ &
 $E_{\mbox{\footnotesize tot}}\!-\!
  E_{\mbox{\footnotesize tot}}^{\mbox{\footnotesize(FSM=0)}}$ (meV) \\
  \hline \hline
  0 & $-2J$ & ~0 \\
  4 &   0   & ~6 \\
  8 & $~6J$ & 60 \\
\hline
\end{tabular*}
\end{center}
\end{table}

\noindent
It should be noted that the AFM configuration, being a broken-symmetry
state, is \emph{not} a faithful representation of the true $S$=0
ground state of the system. Moreover it is obviously a frustrated state,
because each Ni atom interacts with one Ni magnetized in parallel 
and two antiparalelly magnetized counterparts, whereas all coupling
constants are expected to be identical, for symmetry reasons.
Nevertheless, the calculated total energy of this broken symmetry state 
(Table I) may be mapped onto the predictions of the Heisenberg model,
taken together with the results for other two lowest-energy configurations,
according to the FSM analysis of Fig.~\ref{fig:fsm}. 

\section*{5. STRUCTURE RELAXATION}

\begin{figure}[b!] 
  \begin{center}
  \includegraphics*[width=13.0cm]{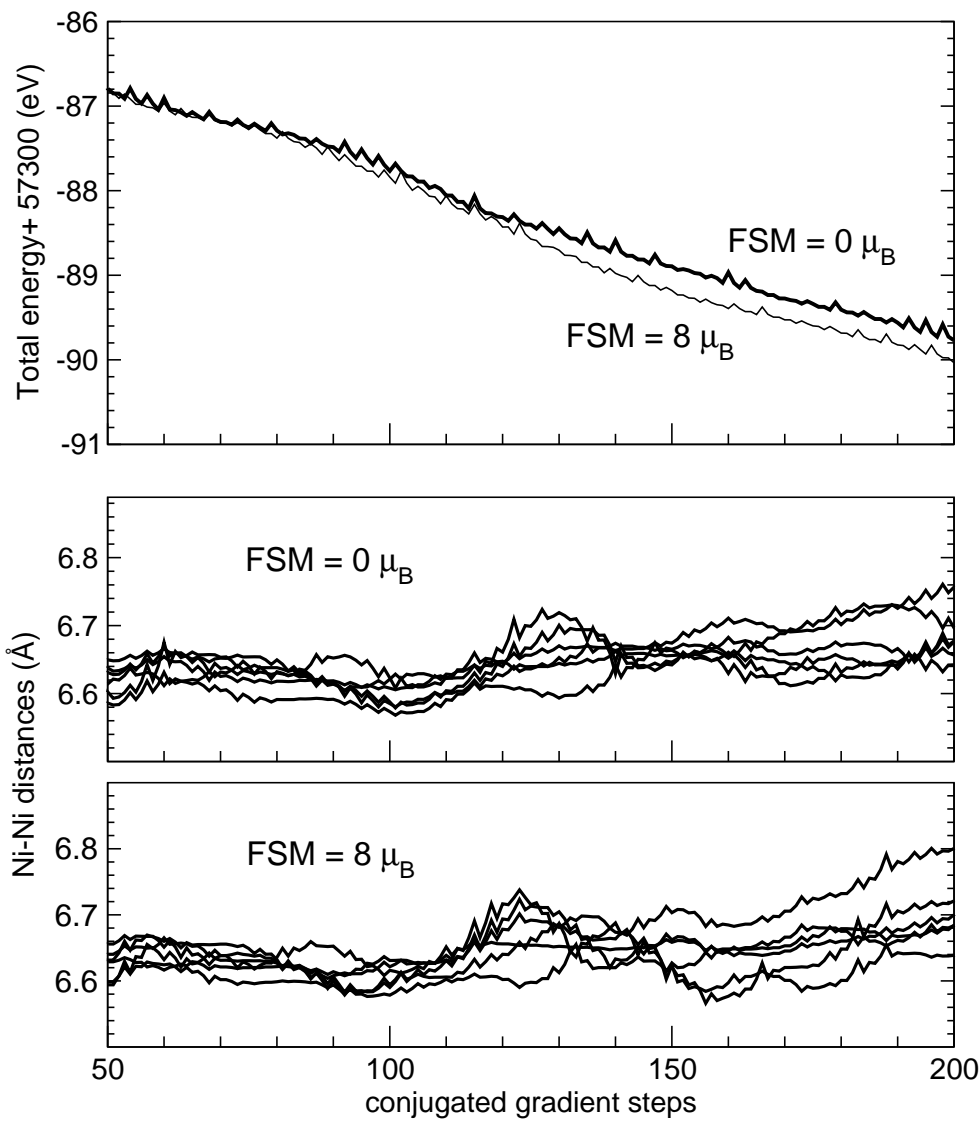}
  \end{center}
\medskip
\caption{\small
Total energy (top panel) and six Ni--Ni distances (two bottom panels)
in the course of unconstrained structure optimization, for
FSM values 0 $\mu_{\mbox{\tiny B}}$ (AFM state) and
8 $\mu_{\mbox{\tiny B}}$ (FM state).
\label{fig:relax}
}
\end{figure}

In conclusion we discuss the structure optimization of the ``Ni$_4$''
molecule in AFM and FM configurations, which correspond to the FSM 
values of 0 and 8 $\mu_{\mbox{\tiny B}}$, correspondingly.
The aim of the simulation was to check whether the interatomic distances 
in the molecule would tend to become pronouncedly different in two 
magnetic configurations, that would probe the assumption of large 
magnetostriction in the molecule. Our preliminary results 
(Fig.~\ref{fig:relax}) does not seem to confirm the latter hypothesis,
but definite conclusion can be only be made on the basis of
fully relativistic calculations.
Our present simulation includes relativistic effects but the spin-orbit 
interaction; consequently, the influence of magnetic structure on total
energy and forces is only via the hybridisation which is slightly
different in FM and AFM configurations. 
When starting from the reported experimental geometry,
an initial stage of relaxation (not shown in Fig.~\ref{fig:relax}) 
substantially lowered the total energy, and induced strong fluctuations 
of interatomic distances. Eventually the molecule
recovers Ni--Ni distances of 6.6--6.7 {\AA}, not pronouncedly different
in FM and AFM cases. A subsequent smooth lowering of the total energy  
does not yet seem converged after 200 conjugated gradient steps.

\noindent
The fact that the total energy of the FM configuration in Fig.~\ref{fig:relax}
lies, over a large number of simulation steps, lower than that of the AFM,
does not contest our above result concerning the system's tendency
for the AFM coupling. The two curves in the upper panel 
of Fig.~\ref{fig:relax} reflect two different relaxation histories,
so that two solutions which accidentally correspond to the same step number
stay in no obvious relation to each other. 
A meaningful comparison of the FM and AFM total energies requires
either the identical structure (as it has been done in the previous
section), or full relaxation of structures in two magnetic configurations
to be compared.

\noindent
Summarizing, we attempted a fit of experimental magnetic susceptibility
and magnetization data for the ``Ni$_4$'' magnetic molecule to the
Heisenberg model, including zero-field splitting (single-ion
anisotropy) and allowing variations of exchange parameters $J$ 
with the external magnetic field.
It seems that fixed (magnetic field and temperature-independent)
values of $J$ do not allow to achieve an acceptable fit of
experimental dependencies. As a microscopic reason for the $J$ dependency
on magnetic field, we assume a magnetostriction in the ``Ni$_4$'' molecule
to play an important role. Without the deformation of the molecule
in the magnetic field taken into account, the results of first-principles
calculations are consistent with experimental data: they indicate
an antiferromagnetic interaction between Ni atoms, and confirm the
conclusion that the magnetic properties of the system cannot be fit
to the isotropic Heisenberg model. An attempt to simulate the relaxation
of the molecule from first principles indicated a not very pronounced
difference in the Ni--Ni distances between ferromagnetic and
antiferromagnetic configurations of the molecule. However, the spin-orbit
interaction has not been included in the present simulation; yet it is
essential for a precise treatment of magnetic anisotropy and magnetostriction,
so its inclusion is likely to affect the results.
\\*[0.5cm]
\noindent {\normalsize \bf \em Acknowledgements} 
\\*[0.5cm]
The authors thank the Deutsche Forschungsgemeinschaft for
financial support (Priority Program SPP 1137 ``Molecular Magnetism'').
AVP acknowledges useful discussions with Paul K\"ogerler 
and Stefan Bl\"ugel.

\end{document}